\documentstyle[eqsecnum,aps,prd,epsf]{revtex}
\date{March 30, 2000}

\tighten
\begin{document}

\newcommand\lsim{\mathrel{\rlap{\lower4pt\hbox{\hskip1pt$\sim$}}
    \raise1pt\hbox{$<$}}}
\newcommand\gsim{\mathrel{\rlap{\lower4pt\hbox{\hskip1pt$\sim$}}
    \raise1pt\hbox{$>$}}}

\title{Extending the Velocity-dependent One-scale String Evolution
Model}

\author{C. J. A. P. Martins \thanks{Also at C. A. U. P.,
Rua das Estrelas s/n, 4150 Porto, Portugal.
Electronic address: C.J.A.P.Martins\,@\,damtp.cam.ac.uk}
and E. P. S. Shellard \thanks{Electronic address:
E.P.S.Shellard\,@\,damtp.cam.ac.uk}
}

\address{Department of Applied Mathematics and Theoretical Physics\\
Centre for Mathematical Sciences, University of Cambridge\\
Wilberforce Road, Cambridge CB3 0WA, U.K.}

\maketitle
\begin{abstract}
We provide a general overview of the velocity-dependent
one-scale model for cosmic string evolution and discuss two further
extensions to it. We introduce and justify a new ansatz for the
momentum parameter $k$, and also incorporate the effect of radiation
backreaction.  We thus discuss the evolution of the basic large-scale
features of cosmic string networks in all relevant cosmological
scenarios, concentrating in particular on the `scaling' solutions
relevant
for each case. In a companion paper, we show, by comparing with
numerical
simulations, that this model provides an accurate description of the
large-scale features of cosmic string networks.
\end{abstract}

%%%%%%%%%%%%%%%%%%%%%%%%%%%%%%%%%%%%%%%%%%%%%%%%%%%%%%%%%%%%%%%%%%%%%%%%%%
\section{Introduction}
\label{sint}

The velocity-dependent one-scale (VOS) model provides the most
convenient
and reliable method by which to calculate the large-scale quantitative
properties of a string network in cosmological and other
contexts\cite{ms1,ms2,thesis}. It is widely used for making
quantitative predict
ions of
the potential
observational implications of cosmic strings\cite{vsh}.
Given its simplicity,
it is remarkable how well the VOS model performs when tested against
high resolution numerical simulations of string networks\cite{test}.
It is well-known that string evolution is a complex physical process
with
a build-up of small-scale structure on the strings,
which is very computationally demanding to model
accurately\cite{at,bb,as}.
Analytic approaches like the VOS model, abandon the possibility
of describing the statistical physics of
the string network accurately and concentrate instead on its
thermodynamics. In other
words, a small number of macroscopic quantities are selected and
the microscopic string equations of motion are used to derive
evolution equations for these averaged quantities.
The price to be paid in this approach is that the averaging process
introduces  phenomenological parameters whose values are not
specified by the model itself. Instead, one must still
fix these parameters by direct comparison with numerical simulations.

The VOS model is a generalization of the `one-scale' model pioneered
by
Kibble\cite{kib} (see also ref.\cite{Ben}) which describes string
motion
in terms of a single correlation length $L$.  By incorporating a
variable
rms velocity $v$, the VOS model extends its
validity into early regimes with frictional damping and
across the important matter-radiation transition, thus giving a
quantitative
picture of the complete  history of a  cosmic string network.
Other analytic approaches to string evolution have attempted to
incorporate the additional small-scale structure seen in numerical
simulations. This includes a `kink-counting' model\cite{AllCal}, a
functional approach \cite{emb}, a `three-scale' model\cite{ack} and a
`wiggly'
model \cite{prep3}.  While these are important for characterising
detailed network
features, they introduce a significant number of further
phenomenological parameters which must be fixed by simulations
(and which remain rather uncertain).  Nevertheless, for describing the
large-scale properties of a long-string network, the VOS model has
proved
to be sufficient for a
good quantitative fit using only a single parameter, the loop
chopping efficiency $\tilde c$.

The purpose of the present paper is, first, to provide a concise
exposition of the VOS string evolution model.  We summarise how it can
be applied to describe cosmological string evolution, including late
times with a cosmological constant, and we present the very different
histories of both GUT- and electroweak-scale strings.  Secondly, we
propose an
improvement of the VOS model by presenting a new ansatz for the
momentum
parameter $k$, which we justify both analytically and numerically.
Thirdly, we present a further extension incorporating radiation
backreaction, which provides small corrections to the cosmological 
scaling laws and which also compares favourably with published results
of
global string simulations.  Finally, we review generalizations in a
curved
FRW spacetime, giving some further asymptotic scaling solutions.
We report on detailed comparisons between
numerical string simulations and the VOS
model elsewhere \cite{test,frac,prep2}.

%%%%%%%%%%%%%%%%%%%%%%%%%%%%%%%%%%%%%%%%%%%%%%%%%%%%%%%%%%%%%%%%%%%%%%%%%%

\section{The VOS string network model}
\label{smod}

The velocity-dependent one-scale model has been described in
considerable detail elsewhere \cite{ms1,ms2,thesis,open,ms3,acm2},
so here we limit ourselves to
a brief summary which highlights the
features that will be important for what follows.  Also for
simplicity, we will
only discuss the evolution of the long string network, even though
this
formalism is also applicable to the loop population, {\it mutatis
mutandis}.
We will discuss this case in detail elsewhere \cite{prep2}.

\subsection{The averaged evolution equations}

Averaged  quantities which we could use to describe the
string network are its energy $E$ and RMS velocity $v$ defined by
\begin{equation}
E=\mu a(\tau)\int\epsilon d\sigma\, , \qquad
v^2=\frac{\int{\dot{\bf x}}^2\epsilon d\sigma}{\int\epsilon
d\sigma}\,,
\label{eee}
\end{equation}
where the string trajectory ${\bf x}(\sigma, t)$ is parametrised by
the
worldsheet coordinates $\sigma$ and $t$ and the `energy density'
$\epsilon
(\sigma,t)$
gives the string length per unit $\sigma$ along the string.

Any string
network divides fairly neatly into two distinct populations, long
(or `infinite') strings and small closed loops with corresponding
quantities denoted by a subscript $\infty$ and $\ell$ respectively.
The long string network
is a Brownian random walk on large scales and can be characterised by
a correlation length $L$.  This can be used to replace the energy
$E_\infty
= \rho_\infty V$
in long strings in our averaged description, that is,
$$
\rho_\infty \equiv {\mu \over L^2}\,.
$$
A phenomenological term must then be included to account for the loss
of
energy from long strings by the production of loops, which are much
smaller
than $L$. A `loop chopping
efficiency' parameter $\tilde c$ is introduced to characterise this
loop production as
\begin{equation}
\left(\frac{d\rho_{\infty}}{dt}\right)_{\rm to\ loops}=
{\tilde c}v_\infty\frac{\rho_{\infty}}{L}
\, . \label{rtl}
\end{equation}
In this approximation,
we would expect the loop parameter $\tilde c$ to remain constant
irrespective of the cosmic regime,
because it is multiplied by factors which determine the string
network self-interaction rate.

From the microscopic string equations of motion, one can then
average to derive the evolution equation for the
correlation length $L$,
\begin{equation}
2\frac{dL}{dt}=2HL(1 + {v_\infty^2})+ {{L}
\over \ell_{\rm f}}{v_\infty^2}
+{\tilde c}v_\infty \, , \label{evl0}
\end{equation}
where $H$ is the Hubble parameter and $\ell_{\rm f}$ is a friction
damping
length scale.  The first term in (\ref{evl0}) is
due to the stretching of the network by the Hubble expansion which is
modulated
by the redshifting of the string velocity.  The second term is due to
frictional interactions by a high density of background particles
scattering
off the strings.  The friction length scale $\ell_{\rm f}$ (defined in
ref.~\cite{ms1}) typically
depends on the background temperature as $\ell_{\rm f} \approx
{\cal o}(1)\mu T^{-3}$, so that it grows with the scale factor as
$a^{3}$.
It usually becomes irrelevant after a time $t_* \approx
(G\mu)^{-1}t_c$ (with
$t_c$ being the epoch at which the network was formed) which
is a very short time for GUT-scale strings but can be as late as $t_0$
for
electroweak strings.
Note also that equation (\ref{evl0}) is valid for an arbitrary flat
FRW model with  $H$ given by
the Friedmann equation,
\begin{equation}
H^2 \equiv \left ( \dot a\over a\right )^2
= H_0^2\left({\Omega_{{\rm R}0}\, a^{-4}}+ {\Omega_{{\rm M}0}\,
a^{-3}}\right)
 + {\textstyle{1\over3}}{\Lambda}\,. \label{Friedmann}
\end{equation}
where $\Omega_{{\rm R}0}$ and $\Omega_{{\rm M}0}$ are the fractional
radiation and matter densities today at $t_0$, $\Lambda$ is the
cosmological constant and we take $a(t_0)=1$.

One can also derive an evolution equation for the long string
velocity with only a little more than Newton's second law
\begin{equation}
\frac{dv_\infty}{dt}=\left(1-{v^2_\infty}\right)
\left[\frac{k}{L}-\left(2H+{1\over \ell_{\rm
f}}\right)v_\infty\right]\, ,
\label{evv0}
\end{equation}
where $k$ is called the `momentum parameter'.  The first term is the
acceleration due to the curvature of the strings and the second
damping
term is from both the expansion and background friction.  The
parameter $k$ is
defined by
\begin{equation}
k\equiv\frac{\langle(1-{\dot {\bf x}^2})({\dot {\bf x}}\cdot {\bf
u})\rangle}
{v(1-v^2)}\, ,
\label{klod}
\end{equation}
with ${\dot {\bf x}}$ the microscopic string velocity and ${\bf u}$
a unit vector parallel to the curvature radius vector.
In previous work \cite{ms1,ms2,ms3,thesis}, we left $k$
as a second phenomenological parameter, while pointing out that it is
related
to small-scale structure and also demonstrating
specified asymptotic dependencies on the velocity. In the next
section, however, we justify an accurate ansatz for $k$ which removes
this
additional freedom.  For most relativistic regimes relevant to cosmic
strings
it is sufficient to define it as follows:
\begin{equation}
k_{\rm rel}(v) =\frac{2\sqrt{2}}{\pi}\;\frac{1-8v^6}{1+8v^6}\, .
\label{krel}
\end{equation}
In the extreme friction-dominated case ($v\rightarrow0$),
we have the nonrelativistic limit
$k_{\rm nr} = 2\sqrt{2}/\pi$ with a more complicated ansatz than
(\ref{krel})
interpolating between these limits for intermediate regimes.

Finally, we end this summary by noting that the VOS model has been
extensively
compared with the results of numerical
simulations\cite{bb,as,test,prep2}
and shown to provide a good fit to the large-scale properties
of a string network.  In particular, it matches well the
evolution between
asymptotic regimes as a network passes through the matter--radiation
transition.  Comparisons with numerical
simulations confirm the constancy of the
only free parameter, the loop chopping efficiency $\tilde c$, and fix
its value to be\cite{test}
\begin{equation}
\tilde c = 0.23 \pm 0.04\,.
\label{cpar}
\end{equation}
The VOS model for any flat FRW cosmology, then, consists of the
evolution equations (\ref{evl0})
and (\ref{evv0}) with the parameters $c$ specified in (\ref{cpar}) and
$k$ given by (\ref{krel}) (or a more accurate general expression given
below) and the scale factor $a$ satisfying the
Friedmann equation (\ref{Friedmann}).

\subsection{Scale-invariant solutions}

We now start to use the VOS model to provide a
general overview of the evolution of string networks in various
cosmological scenarios.  First we analyse some basic late-time
properties of cosmic string networks, neglecting
the effect of friction due to particle scattering.
A crucial question is
whether or not they can reach a `scale invariant' attractor
solution which is required, among other things, for a
Harrison-Zel'dovich
spectrum of primordial density fluctuations to be generated. This can
be
discussed by analysing the VOS equations
(\ref{evl0}) and (\ref{evv0}).

Scale-invariant solutions of the form $L\propto t$, $L\propto H^{-1}$
or $L\propto d_H$, together with $v_\infty=const.$, only appear to
exist when
the scale factor is a power law of the form
\begin{equation}
a(t)\propto t^\beta\, , \qquad \beta=const.\, , \qquad 0<\beta <1\,
\,.
\label{conda0}
\end{equation}
This condition implies that
\begin{equation}
L\propto t\propto H^{-1}\propto d_H\, ,
\label{props}
\end{equation}
with the proportionality factors dependent on $\beta$.
It is useful to introduce the following useful
parameters to describe the relative correlation length and densities,
defining them
respectively as
\begin{equation}
L=\gamma t\,,  \qquad
\zeta \equiv \gamma^{-2}= \rho_\infty t^2/\mu\,.
\end{equation}
By looking for stable fixed points in the VOS equations, we can
express the actual scaling
solutions in the following implicit form:
\begin{equation}
\gamma^2=\frac{k(k+{\tilde c})}{4\beta(1-\beta)}\, ,\qquad
v^2=\frac{k(1-\beta)}{\beta(k+{\tilde c})}\, ,
\label{scalsol}
\end{equation}
where $k$ is the constant value of $k(v)$ given
by solving the second (implicit) equation for the velocity.
Although it may not be obvious by inspection, it is easy
to verify numerically that this solution is well-behaved and stable
for
all realistic parameter values.

\begin{figure}
\vbox{\centerline{
\epsfxsize=0.7\hsize\epsfbox{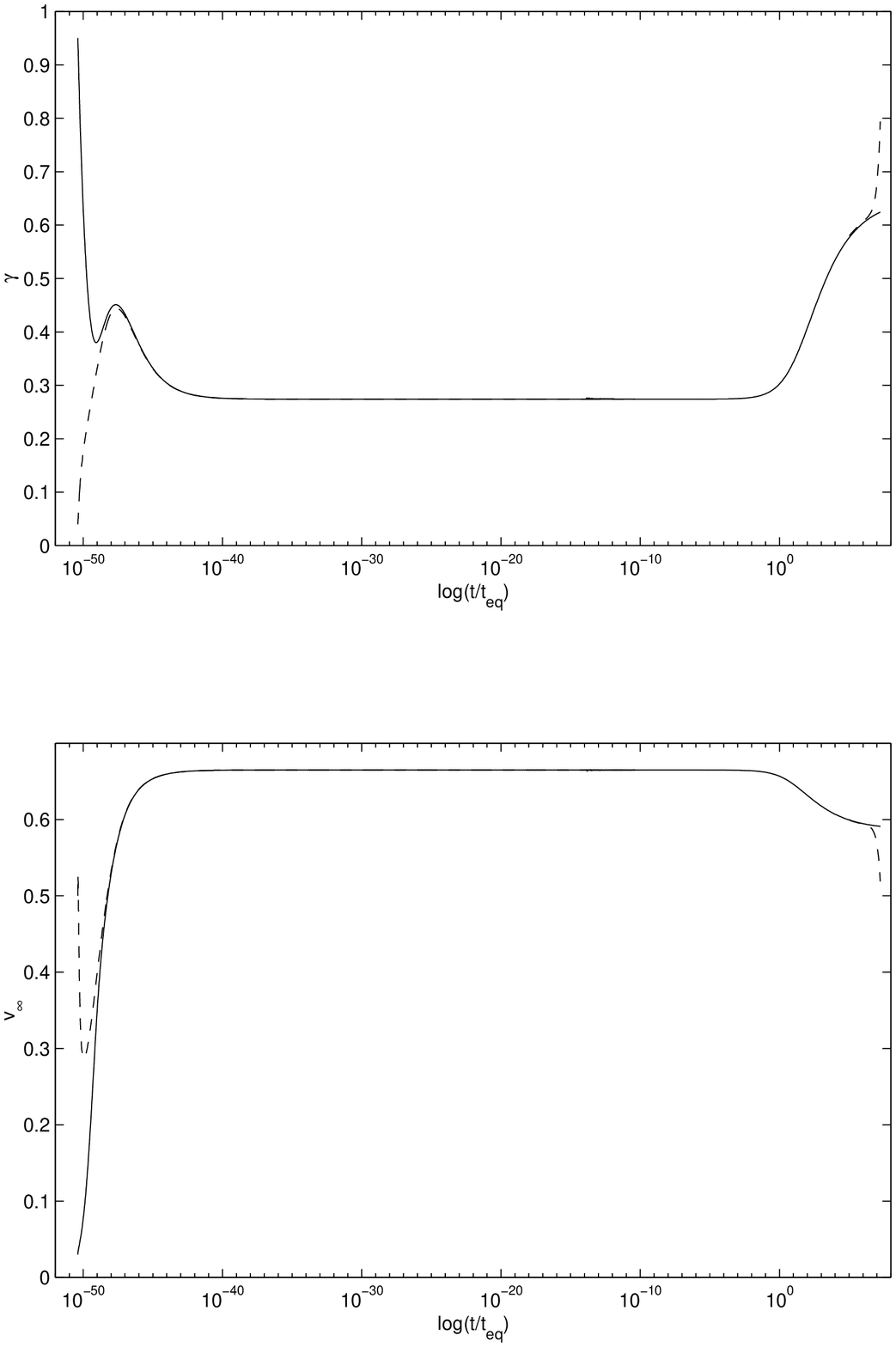}}
\vskip.4in}
\caption{The complete cosmological history of a GUT-scale cosmic
string
network. The evolution of $\gamma\equiv L/t$ is shown in the top
panel,
while the bottom one shows the network's RMS velocity $v_\infty$.
Time is plotted relative to the epoch of equal matter and radiation
densities;
the plots start at the epoch of string formation and end at the
present
day. At early times, the solid curve corresponds to initial conditions
typical of a first-order phase transition, while the dashed one
corresponds
to a second-order transition. At late times the solid curve
corresponds to
a model with $\Omega_m=1$, while the dashed one is for the
observationally
preferred case $\Omega_m=0.2, \Omega_\Lambda=0.8$ (to be discussed
below). Note the deviations to the scaling behaviour.}
\label{fig5}
\end{figure}

If the scale factor is not a power law, then simple scale-invariant
solutions like (\ref{scalsol}) do not exist. Physically this happens
because the network
dynamics are unable to adapt rapidly enough to the changes in
the background cosmology. A prime example of this is, of course, the
transition between the radiation and matter-dominated eras.
The evolution of a GUT-scale network shown in fig. \ref{fig5}
illustrates asymptotic regimes in which string
evolution is
scale-invariant, as well as the matter--radiation transition where it
is not.
Note that for realistic cosmological parameter choices, a string
network today is still only slowly approaching its asymptotic matter
density.
Since, the
cosmological importance of the changes in the network properties
during
the matter--radiation transition cannot be over-emphasised, it is
important to calculate these
accurately either with direct numerical simulations or with the VOS or
similar
analytic model.  The same is also true for late time curvature or
cosmological
constant domination.

\subsection{Friction-dominated scaling solutions}

During friction-dominated epochs one has different `scaling'
solutions.
However, these are no longer `scale-invariant', since in this case the
network retains a memory of its initial conditions, and in particular
of the
epoch of formation. This can be conveniently expressed by a parameter
\begin{equation}
\theta\sim\left(\frac{t_c}{t_{Pl}}\right)^{1/2}\, ,
\label{partheta}
\end{equation}
which is essentially the value of the ratio of the damping terms due
to friction
and Hubble damping, measured at the epoch of string formation.

Thus in realistic cosmological contexts one can have two
different regimes \cite{ms1,ms2,thesis}.
The first is a `stretching' regime,
\begin{equation}
\frac{L}{L_c}=\left(\frac{t}{t_c}\right)^{1/2}\, ,  \qquad
v=\frac{t}{\theta L_c}\, ,
\label{stretchr}
\end{equation}
which is an early-time, transient period which will occur when the
initial
string density and velocity are sufficiently low---for example,
as a result of a slow first-order phase transition.
In this case the network
starts out with a correlation length significantly larger than the
damping
length and so is `frozen', and is conformally stretched. However the
damping length is growing as $\ell_f\propto t^{3/2}$, so it quickly
catches
up with it, ending this regime. However, this can last for many orders
of
magnitude in time for electroweak-scale networks. Although this is not
cosmologically relevant except for extremely light strings, 
the analogous regime in the matter-dominated case
would be
\begin{equation}
L\propto t^{2/3}\, ,  \qquad
v\propto t^{4/3}\, .
\label{stretchm}
\end{equation}

The true attractor solution for a friction-dominated epoch, which
follows the
stretching regime (if this exists) is the Kibble regime, which in the
radiation era has the form
\begin{equation}
\frac{L}{L_c}=\left[\frac{2k_{nr}({\tilde
c}+k_{nr})}{3\theta}\right]^{1/2}\left
(\frac{t}{t_c}\right)^{5/4}\, ,  \qquad
v=\left[\frac{3k_{nr}}{2\theta({\tilde
c}+k_{nr})}\right]^{1/2}\left(\frac{t}{t_
c}\right)^{1/4}\, ,
\label{Kibrad}
\end{equation}
where $k_{nr}$ is the value of the momentum parameter in the
nonrelativistic
limit given above.
In this case the correlation length stays halfway 
between the damping length and the horizon length.
Again there is a matter era analogue, which as the form
\begin{equation}
L\propto t^{3/2}\, ,  \qquad
v\propto t^{1/2}\, ,
\label{Kibmat}
\end{equation}
but this is rarely relevant cosmologically. In the extreme case of
an electroweak scale network, friction domination ends after
radiation-matter
equality, but not far enough away from it for this regime to be
reached.

\begin{figure}
\vbox{\centerline{
\epsfxsize=0.7\hsize\epsfbox{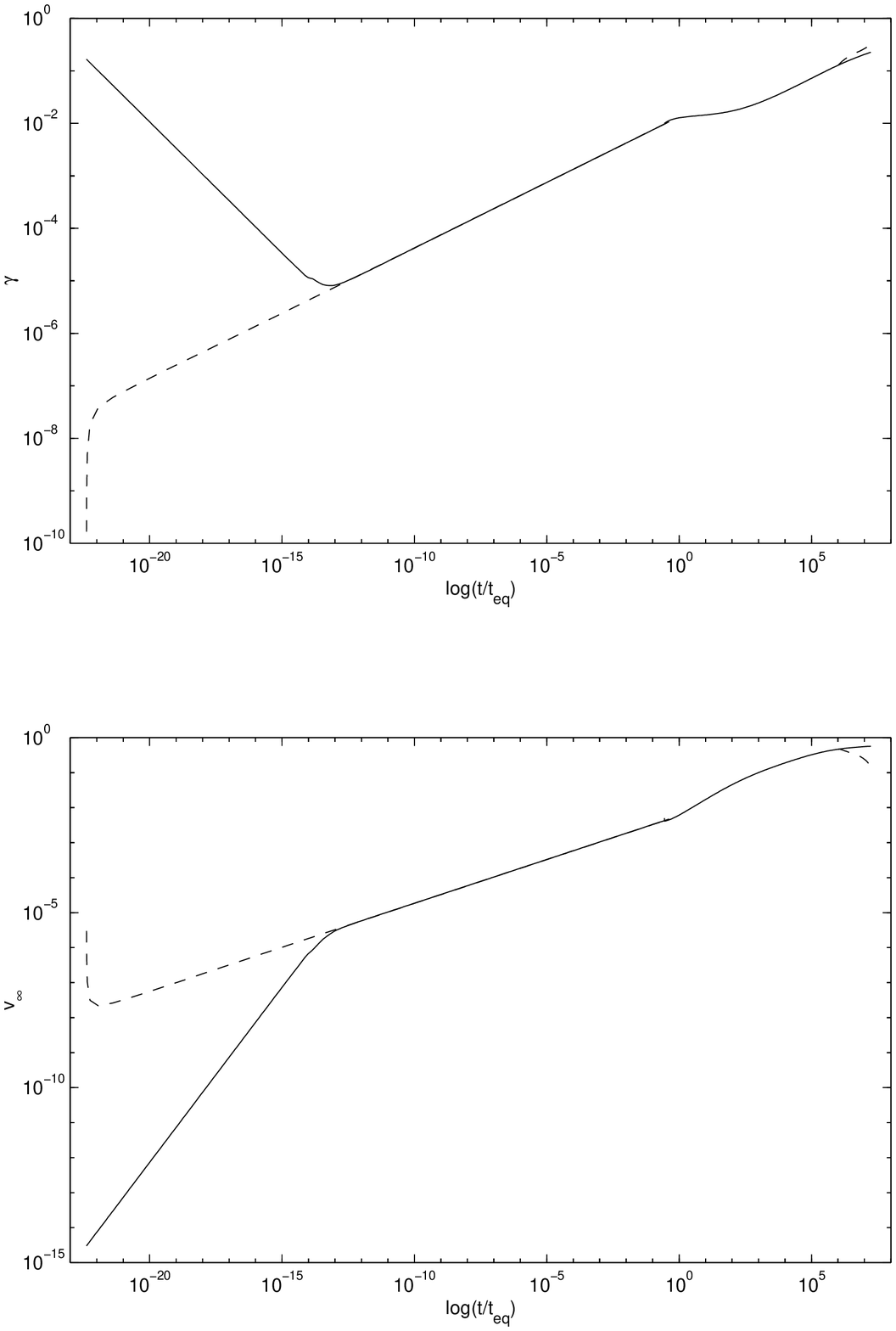}}
\vskip.4in}
\caption{Same as Fig. \ref{fig5}, but for an electroweak-scale cosmic
string network.}
\label{fig6}
\end{figure}

These points are illustrated in fig. \ref{fig6} where we see
the evolution of an
electroweak-scale string network.  It is interesting to note the
scaling
behaviour for two
cases with very different initial conditions from extreme first- and
second-order
transitions.  The high density strings from a second-order transition
quickly approach the Kibble scaling regime (\ref{Kibrad}) discussed
above.
However,
the low density strings from a first-order transition begin in a
distinct stretching regime (\ref{stretchr}) and persist in it with
their
density falling slowly until
it matches that for the attractor
Kibble regime. In this case, the string network retains
a `memory' of its initial density for about ten orders of magnitude
in cosmic time.  However, even during the
friction-dominated regime the network is able to
erase this memory once the Kibble regime is reached. On the other
hand,
the memory of the epoch of formation is not erased---one could in
principle recover it by measuring the parameter $\theta$.
 This can  also be seen for GUT strings in fig. \ref{fig5},
although the
network does not have time to relax into a definite scaling regime
before
friction-domination ends.  For electroweak strings, there are also
interesting
departures from scaling behaviour at the matter--radiation
transition and the network remains friction-dominated until about
three orders
of magnitude in time afterwards. Also note that in the $\Omega_m=1$
case
the strings only reach the relativistic regime at about the present
time,
and in the observationally preferred case $\Omega_m=0.2,
\Omega_\Lambda=0.8$
they are always non-relativistic. This point is crucial, among other
things,
for a quantitative analysis of the evolution of superconducting
strings
and vortons \cite{vortons1,vortons2}.

\subsection{A cosmological constant}

\begin{figure}
\vbox{\centerline{
\epsfxsize=0.9\hsize\epsfbox{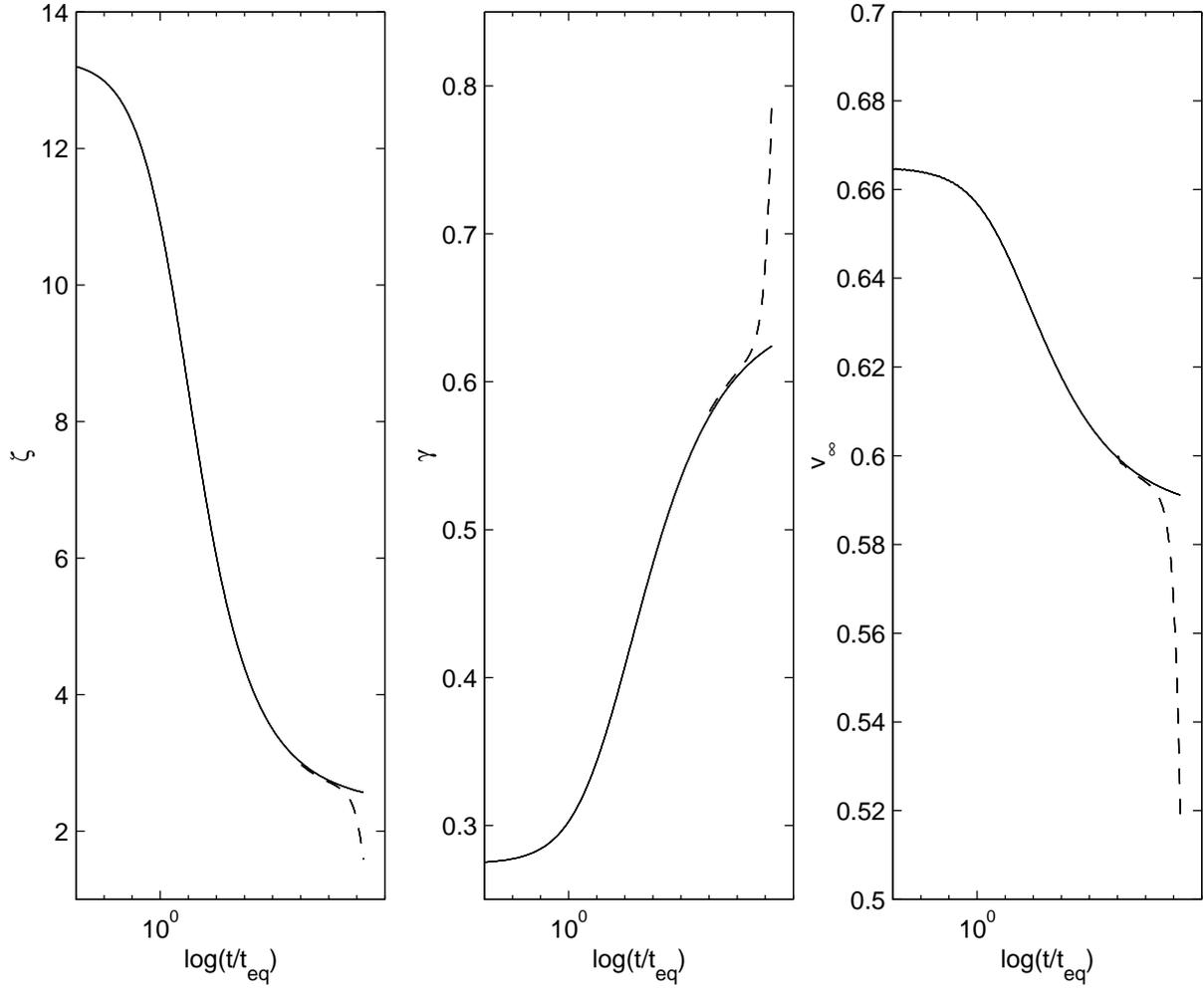}}
\vskip.4in}
\caption{A close-upof Fig. \ref{fig5}, showing the recent evolution of
the
GUT-scale string network. Also shown is parameter $\zeta$, a measure
of the
long-string density. Notice the dramatic chages once the cosmological
constant dominates.}
\label{fig8}
\end{figure}

We can also use the VOS model in a flat background to discuss the
domination at late times by a cosmological
constant $\Lambda$ \cite{carsten,acm2}, a model for which there
appears
to be growing observational evidence. In the extreme asymptotic case
when the universe is inflating we have $a\propto \exp{(Ht)}$ with
$H=\sqrt{\Lambda/3}$.  The network will `freeze out' and will simply
be
conformally stretched, that is,
\begin{equation}
L\propto a\, ,\qquad v_\infty\propto a^{-1}\, ,
\label{scalinf}
\end{equation}
where, as soon as the strings become nonrelativistic $k_{\rm
nr}=2\sqrt{2}/\pi$,
their product satisfies
\begin{equation}
Lv_\infty=\frac{2\sqrt{2}}{\pi}H\, .
\label{condinf}
\end{equation}
Of course, this solution will only apply at early times actually
during inflation.
At the present time we will only be slowly approaching a new
stretching regime,
so we have to solve the VOS model explicitly.  This is shown for a
model in
which $\Omega_\Lambda = 0.8$, as
a dashed line
at late times in figs. \ref{fig5}--\ref{fig6},
as well as in detail for GUT-scale
strings in fig. \ref{fig8}. It is clear
that there is a significant fall in the string density and velocity,
an
effect which, for
example, would affect the large-angle anisotropies in the cosmic
microwave
sky. The evolution of the string network is clearly not
scale-invariant during
any period after $t_{\rm eq}$.

\section{The momentum parameter}

Having introduced the VOS model and some of its key cosmological
implications, we now turn to a more detailed discussion of the
so-called `momentum parameter' $k$ defined in (\ref{klod})
and which is important for solving (\ref{evv0}).  String velocities
are determined by the current acceleration to which they are subjected
from the local string curvature, as well as their `bulk' momentum 
left-over from previous accelerations.  Heuristically,
we can imagine separating the velocity into these the curvature `c' 
and bulk `p' contributions as
${\dot {\bf x}}={\dot {\bf x}}_{\rm c}+
{\dot {\bf x}}_{\rm p}$.
In the
extreme friction-dominated limit, the velocity is entirely due to the
curvature and reaches a limiting average velocity set by
the friction length scale, $v_{\rm c} \equiv \langle{\dot {\bf x}}_{\rm
c}^2 \rangle^{1/2} = \ell_f/L$.  However, as the velocity increases towards
relativistic values we can expect the momentum contribution to become
larger.  Let us suppose that their relative contribution is
proportional
to some power law of the total velocity $v$, that is, $v_{\rm
p}/v_{\rm c}
\propto v^\alpha$ where $\alpha$ is clearly greater than unity.
In this case, one
can find after some straightforward (although tedious) manipulations
that the following approximate relation holds for the momentum
parameter $k$
\begin{equation}
k\sim\frac{1-2^\alpha v^{2\alpha}}{1+2^\alpha v^{2\alpha}}\, .
\label{knew}
\end{equation}
In this we have also used the fact that flat spacetime analytic
calculations \cite{ms2,thesis} have shown that
\begin{equation}
k(1/\sqrt{2})=0\, ,
\label{kflat}
\end{equation}
which holds exactly. Note that the above expression means that
(\ref{knew})
can also be approximately written as
\begin{equation}
k\sim\frac{1-(v_p/v_c)^2}{1+(v_p/v_c)^2}\, .
\label{knew2}
\end{equation}

\begin{figure}
\vbox{\centerline{
\epsfxsize=0.6\hsize\epsfbox{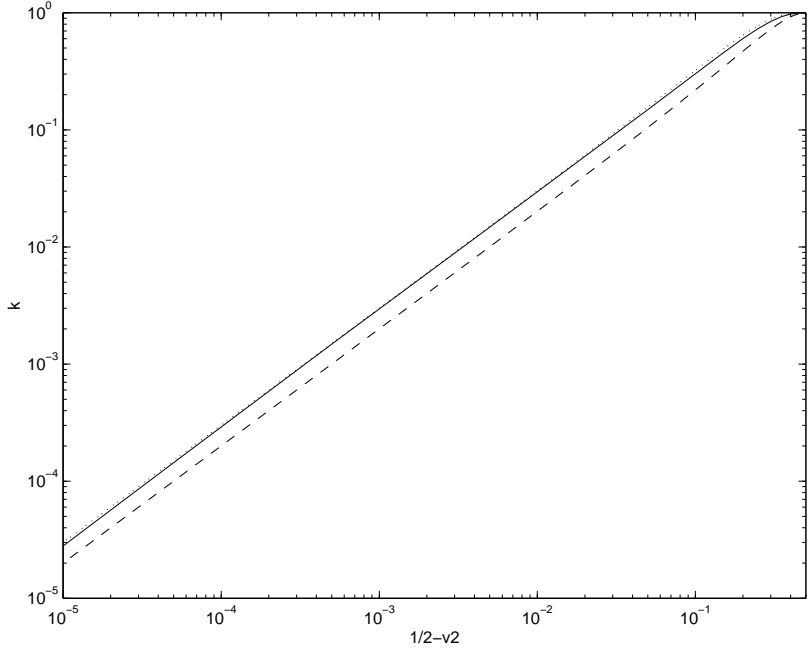}}
\vskip.4in}
\caption{Testing the {\em ansatz} (\ref{knew}) for $k(v)$
in the relativistic limit.
The solid line corresponds to our numerical calculation of k for the
helicoidal string solution, while the dashed and dotted lines
correspond
to the ansatz (\ref{knew}) with $\alpha=2$ and $\alpha=3$
respectively.}
\label{fig1}
\end{figure}

We can determine $\alpha$ by studying the well-known \cite{helix}
helicoidal string solution in
flat space but perturbed by a frictional
force. This solution is 
\begin{equation}
{\bf x}=\left(A\sin{\sigma}(\cos{t}+\eta),
A\cos{\sigma}(\cos{t}+\eta),
\sqrt{1-A^2}\sigma\right)\, ,
\label{helixans}
\end{equation}
where $0\le A\le1$ and $\eta$ is a small perturbation, which vanishes
if there is no friction.  Here,  $A=1$ corresponds to 
a circular loop, while $A=0$ is a static straight string. The
evolution
equation for the perturbation $\eta$ has the form
\begin{equation}
{\ddot \eta}+2{\dot \eta}\frac{\sin{t}\cos{t}}{1-A^2\sin^2{t}}+
\eta\frac{(1-A^2
)\sin^2{t}-\cos^2{t}}{1-A^2\sin^2{t}}=
\frac{\sin{t}}{\ell_f}(1-A^2\sin^2{t})\,
,
\label{perturb}
\end{equation}
where $\ell_f$ is the friction length scale. This can then be solved
numerically,
and from this solution one can calculate $k$. By changing parameters
$A$ and
$\ell_f$ one can do this for a wide range of velocities, and hence
obtain a
plot of $k=k(v)$. This is plotted in Fig. \ref{fig1} and compared with
cases
$\alpha=2$ and $\alpha=3$ of our {\em ansatz} (\ref{knew}). It can be
seen that $\alpha=3$ provides an excellent approximation in this
regime.

There is, however, one problem with this simple {\em ansatz}, namely
that it would
give $k(0)=1$. Even though one might naively expect this to be the
correct
limit, it is not so. This can be seen easily as follows. Assume that
the
velocity of say a loop is determined only by curvature, that is,
neglect the
momentum contribution. (This should be valid in the
non-relativistic case.) Then $k$ will
be approximately given by
\begin{equation}
k\approx\frac{<|{\dot {\bf x}^2}|(1-{\dot {\bf x}^2})>}{v(1-v^2)}\, .
\label{knrlimit}
\end{equation}
This quantity can be easily calculated for the analytic helicoidal
solution,
yielding
\begin{equation}
k=\frac{2\sqrt{2}}{\pi}\frac{1-2A^2/3}{1-A^2/2}\, .
\label{solkaa}
\end{equation}
Hence we find in the small amplitude limit $A\to0$ that
\begin{equation}
k_{\rm nr}=\frac{2\sqrt{2}}{\pi}\approx 0.9\,.
\label{limsolknr}
\end{equation}
We also note in passing that in the relativistic
limit $A\to1$ this same calculation would
give $k\approx 4\sqrt{2}(3/\pi)\approx0.6$ instead of the true value
$k=0$,
which clearly demonstrates the importance of momentum in the
relativistic case.

\begin{figure}
\vbox{\centerline{
\epsfxsize=0.6\hsize\epsfbox{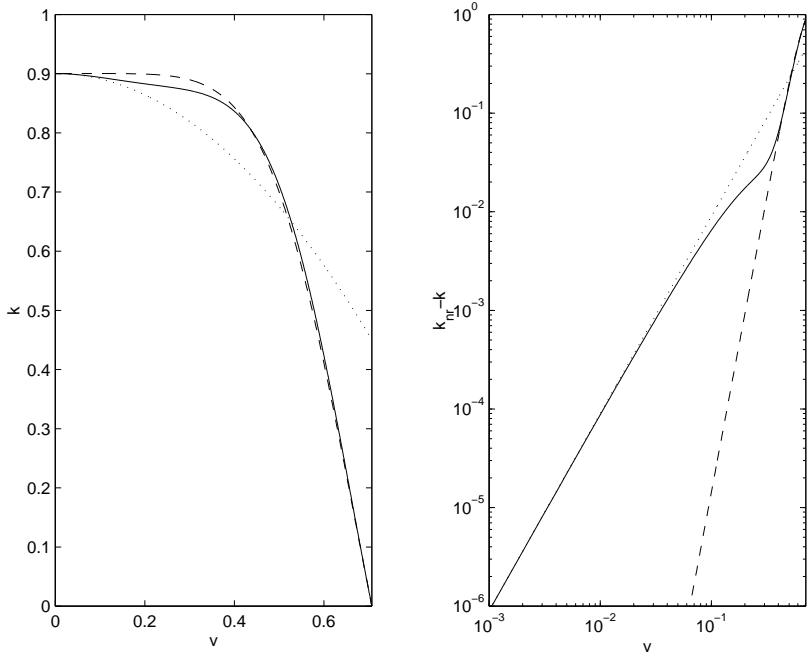}}
\vskip.4in}
\caption{Comparing our full ansatz for the momentum parameter
(\ref{newknew})
(solid line), and the simpler expressions for the relativistic and
friction
dominated regimes, (\ref{krel2}) and (\ref{nonrellim}) (dashed and
dotted,
respectively), for relativistic (left panel) and non-relativistic
(right
panel) velocities}
\label{fig9}
\end{figure}

The final issue to be considered is the transition between the two
regimes. The only reliable way of studying this issue is through
direct measurement in a string network simulations
with ultra-high resolution. We shall report on the
details of this elsewhere \cite{prep2}. Here we simply point out that
we
do confirm the value $k_{\rm nr}$ as the non-relativistic limit. It is
then
easy to find a fitting function for the transition
between the regimes which has the correct asymptotic limits described
previously,
that is, 
\begin{equation}
k(v)=\frac{2\sqrt{2}}{\pi}(1-v^2)(1+2\sqrt{2}v^3)\frac{1-8v^6}{1+8v^6}\,
.
\label{newknew}
\end{equation}
The additional factors are required to reproduce both the relativistic
and
non-relativistic limits accurately. Note that if one is only
interested in the relativistic regime (say for GUT-scale cosmic
strings,
as in the present paper) then the simpler expression (\ref{krel}),
that is
\begin{equation}
k_{\rm rel}(v) =\frac{2\sqrt{2}}{\pi}\;\frac{1-8v^6}{1+8v^6}\, ,
\label{krel2}
\end{equation}
should be sufficiently accurate to provide reliable results. On the
other
hand, a reliable approximation for small non-relativistic velocities
in the 
friction-dominated limit extends 
(\ref{limsolknr}) as 
\begin{equation}
k_{\rm nr}(v)=\frac{2\sqrt{2}}{\pi}(1-v^2)\, .
\label{nonrellim}
\end{equation}
We plot these three ansatze in fig. \ref{fig9}, and also confirm the
validity
of $k_{\rm rel}$ and $k_{\rm nr}$ in the appropriate limits.

Before we end this section, however, it is wise to discuss the
interpretation
of parameter $k$ in this model. It should be kept in mind that this
is,
{\em ab initio}, a {\em phenomenological} parameter, which accounts
for a number
of non-trivial effects related to the presence of small-scale
structures
on the strings. By construction, our model does not explicitly account
for these small-scale effects, and hence they end up somehow encoded
in $k$. One should not therefore infer too much from the aesthetic
qualities of the function $k(v)$ we find---it is simply a
phenomenological parameter that does a good job.
Presumably this parameter will
have a much clearer interpretation in the context of a proper
wiggly string evolution model \cite{carter,prep3}.

%%%%%%%%%%%%%%%%%%%%%%%%%%%%%%%%%%%%%%%%%%%%%%%%%%%%%%%%%%%%%%%%%%%%%%%%%%

\section{The effect of radiation back-reaction}

We now turn to some further extensions of the VOS model.
The effect of gravitational back-reaction on the long-string
network \cite{AllCal,ack} can be included in the evolution equation
for the correlation length (\ref{evl0}) in the same way as
previously achieved for the evolution of the length of a string
loop\cite{ms2}.
For gravitational radiation the following term can be added
to the right-hand side of (\ref{evl0})
\begin{equation}
2\left(\frac{dL}{dt}\right)_{\rm gr}\equiv8\Sigma_{\rm
gr}v^6_\infty=8{\tilde \Gamma}G\mu
 v^6_\infty \, . \label{grb}
\end{equation}
Here, ${\tilde \Gamma}$ is a constant which is a long-string analogue
of the $\Gamma \approx 65$ found for the radiative decay of strings.
Of course, ${\tilde \Gamma}$ will be affected by a number of physical
factors such as the presence of small-scale structure, but we can
expect it to satisfy ${\tilde \Gamma}\lsim \Gamma$ and it would be
surprising if it were very much smaller.  Clearly, due to the high
velocity power ($v^6$) involved in radiative backreaction,
this term will not be important in any regime where string motion is
strongly friction-dominated (and hence non-relativistic). We note also
that there is an interesting coincidence in the ansatz (\ref{krel})
for the momentum parameter $k$ which
too has a $v^6$ power, but we are unsure as yet whether this has any
deeper
significance.

For global string radiation
into Goldstone bosons or axions,
the corresponding radiative decay term at a time $t$ will be
\begin{equation}
2\left(\frac{dL}{dt}\right)_{\rm ax}\equiv8\Sigma_{\rm
ax}v^6_\infty=\frac{8{\tilde \Gamma} v^6_\infty}{
2\pi \ln (t/\delta)}  \, , \label{ax}
\end{equation}
where the logarithmic term arises because of the long-range fields
of the global string and $\delta$ is the string width.  For
cosmological
GUT-scale strings, the backreaction term for local strings
is $\Gamma G\mu\sim 10^{-4}$
whereas for global strings it is about three orders of magnitude
larger.

Note that the velocity equation has  no correction at this order
due to the gravitational back-reaction effects.
Such effects are already included through the string
curvature, which acts as a source for the velocity equation (i.e., the
$1/L$
term), which will be different in this case.

Remarkably, the inclusion of the back-reaction term does not affect
the
existence of a scale-invariant attractor solution. However,
it does of course influence the quantitative values of the scaling
parameters,
as well as the timescale necessary for this
solution to be reached. For example, the inclusion of back-reaction
can make the approach to scaling much faster.

If the gravitational back-reaction is non-zero, one can distinguish
two asymptotic
cases.
Firstly, if $\Sigma$ is small (of order unity at most) then
the effect of back-reaction on the scaling solution will also be
small.
This will be the case, for example, for most local or global string
networks in
a cosmological context.
We can express this as
\begin{equation}
\gamma^2\approx\gamma_0^2\left(1+\Delta\right)\, ,
\label{scalgammasmallg}
\qquad v^2\approx v_0^2\left(1-\Delta\right)\, ,
\label{scalvelsmallg}
\end{equation}
where $\gamma_0$ and $v_0$ are the ``unperturbed'' scaling
values, given by eqns. (\ref{scalsol}),
and the back-reaction correction has the form
\begin{equation}
\Delta=8\beta v_0^5\Sigma=8\beta
\left[\frac{k(1-\beta)}{\beta(k+{\tilde c})}\right]^{5/2}\Sigma\, .
\label{vecorrect}
\end{equation}

On the other hand, for large enough values of $\Sigma$, the
back-reaction term will dominate the evolution equation for the string
length scale $L$, and the attractor scale-invariant solution has a
different
form altogether. It is not possible to write this solution in closed
form,
even expressing $k$ implicitly as above. However, it is possible to
write
it as a series. The dominant term and the first correction take the
form
\begin{equation}
\gamma=\frac{k}{2\beta}\left[\frac{8\beta\Sigma}{k(1-\beta)}\right]^{1/7}\left(1+
\Delta_1+\ldots\right)\, ,
\label{scalgammabigg}
\end{equation}
\begin{equation}
v=\left[\frac{k(1-\beta)}{8\beta\Sigma}\right]^{1/7}\left(1-\Delta_1+\ldots\right
)\, ,
\label{scalvelbigg}
\end{equation}
with
\begin{equation}
\Delta_1=\frac{1}{2^{6/7}7}(k+{\tilde
c})\left[\frac{\beta}{k(1-\beta)}\right]^{5/7}
\Sigma^{-2/7}\, .
\label{vecorrect11}
\end{equation}

\begin{figure}
\vbox{\centerline{
\epsfxsize=0.7\hsize\epsfbox{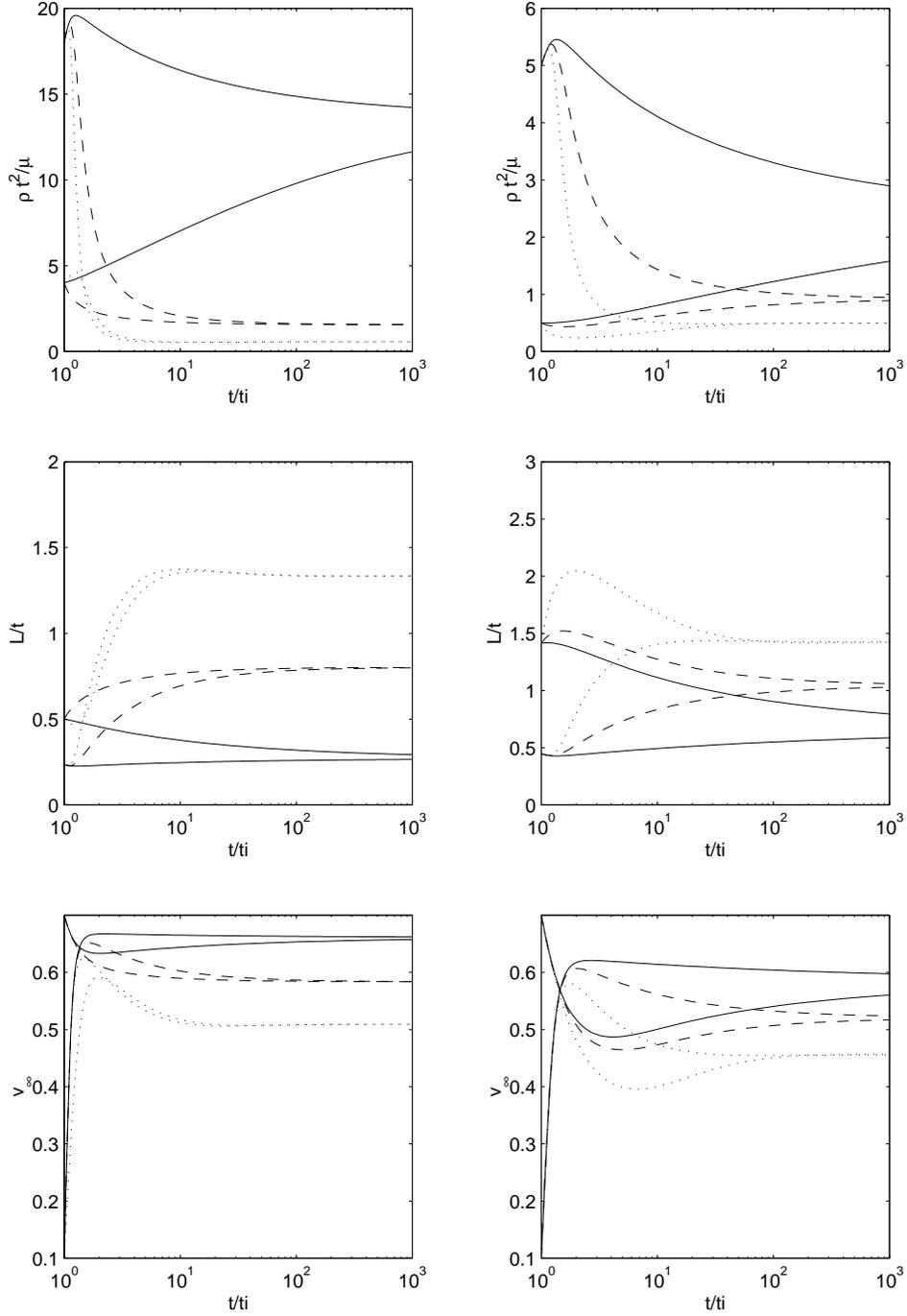}}
\vskip.4in}
\caption{The effect of gravitational back-reaction on the approach to
scaling
of a GUT-scale cosmic string network. Left-side panels correspond to
the
radiation era, while right-side ones are for the matter era. All plots
have ${\tilde c}=0.23$; the back-reaction parameter is respectively
$\Sigma=0$ (solid curves), $\Sigma=10$ (dashed)
and $\Sigma=50$ (dotted). For each case two curves are plotted,
corresponding to initial conditions typical of a first-order (low
density and velocity) or second-order (high density and velocity)
phase transition. The effect of friction due to particle scattering
has been
neglected, in order to mimic currently existing numerical
simulations.}
\label{fig7}
\end{figure}

In fig.~\ref{fig7} we plot the approach to scaling of some relevant
string networks in the radiation and matter eras. 
The different timescales for convergence
towards the attractor solution are clearly noticeable.
Here, we have chosen
initial conditions that would correspond to somewhat extreme first and
second
order phase transitions.  Also note that for the radiation
era we have neglected the effect of friction due to particle
scattering, in
order to reproduce the initial conditions often used in numerical
simulations
of string networks.
For each of the cases above, three curves are plotted, corresponding
to
the values $\Sigma=0$ (no back-reaction), $\Sigma=1.25$
(close to the maximum value that can be accurately described by the
scaling solution (\ref{scalgammasmallg}--\ref{vecorrect})) and
$\Sigma=5.5$ (beyond which the scaling
solution (\ref{scalgammabigg}--\ref{vecorrect11}) becomes accurate).

For large $\Sigma$, the effects of back-reaction seen in
fig.~\ref{fig7} on
could be quite dramatic for the string network density, but note that
they are much less drastic 
for the string velocities. In particular, we emphasize
that gravitational back-reaction alone does not slow down a string
network to non-relativistic speeds---only a friction-dominated regime
can achieve this.

Interestingly, there has been recent work 
on numerical simulations of global string networks \cite{yamaguchi}
which explore the strong backreaction regime described by 
(\ref{scalgammabigg}--\ref{vecorrect11}). 
These authors report 
a surprisingly  low string density relative to the gauged case. 
For their expanding universe simulations in the realistic case 
with periodic boundary conditions, they find the following radiation 
and matter era densities respectively, 
\begin{equation}
\zeta_{\rm rad}=0.9\pm0.1\,, \qquad
\zeta_{\rm mat}=0.5\pm0.1\, .
\label{yama2}
\end{equation}
These results are perfectly consistent (within the estimated error
bars)
with our extended VOS model if we adopt a back-reaction parameter
\begin{equation}
\Sigma_{\rm ax\hbox{-}sim}\approx 3\,.
\label{grbrglobal}
\end{equation}
Indeed, this corresponds to 
the approximate average 
value for $\Sigma_{\rm ax}$ that one would estimate for simulations
of this resolution. 
Present limitations on numerical dynamic range give the 
upper bound $\ln (t/\delta)\lesssim 5$ (at the end of the simulation),
implying $\Sigma_{\rm ax\hbox{-}sim}\gtrsim 2$ throughout. 

It is important to note, however, that the immediate extrapolation of
these 
results (\ref{yama2}) to a cosmological context would be erroneous.
For
example, for GUT-scale global strings, at the present time we can
expect 
$\ln (t/\delta)\gtrsim 100$, which implies the appropriate
backreaction 
parameter in this case will be $\Sigma_{\rm ax}\lesssim 0.1$.  Such
cosmic
global strings are firmly in the regime
(\ref{scalgammasmallg}--\ref{vecorrect})
where backreaction effects are small and, in this case, these would
reduce
the local string density by less than 10\%.  We make a more detailed 
comparison with \cite{yamaguchi} elsewhere \cite{test}.

%%%%%%%%%%%%%%%%%%%%%%%%%%%%%%%%%%%%%%%%%%%%%%%%%%%%%%%%%%%%%%%%%%%%%%%%%%
\section{String networks in general FRW spacetimes}

In this section we discuss the behaviour of our model in more general
FRW universes, and in particular in open
universes \cite{open,thesis,ms3,carsten}.

The evolution equations will obviously be affected by the different
behaviour
of the scale factor, as given by the Friedmann equation
\begin{equation}
H^2 \equiv \left ( \dot a\over a\right )
= H_0^2\left({\Omega_{{\rm R}0}\, a^{-4}}+ {\Omega_{{\rm M}0}\,
a^{-3}}+ {\Omega
_{{\rm Q}0}\, a^{-m}}\right)
 + {\textstyle{1\over3}}{\Lambda}-Ka^{-2}\,. \label{Friedmann2}
\end{equation}
Note that we are allowing for curvature, and also for an extra fluid
whose energy density decays as $a^{-m}$. It should be kept in mind
that
the Friedman equation should, in general,
contain a contribution for the string density, since it is possible
that
this becomes cosmologically important.

However, apart from these
effects, one must also include
an additional correction due to the curvature \cite{ms3,thesis}.
One should note that this is essentially the
curvature radius of the strings, $L$,
divided by the radius of spatial curvature of the universe,
\begin{equation}
{\cal R}=\frac{H^{-1}}{\mid1-\Omega\mid^{1/2}}\, . \label{curvrad}
\end{equation}
Indeed, after a certain amount of algebra, one finds correction terms
that are of the form
\begin{equation}
w=1-(1-\Omega)(HL)^2\, . \label{esse}
\end{equation}
Note that $\Omega$ denotes the total density of the universe.
For a universe with a critical density, $\Omega=1$, we have $w=1$.

The evolution equation for the
correlation length $L$ now takes the form
\begin{equation}
2\frac{dL}{dt}=2HL+\frac{L}{\ell_d}\frac{v_\infty^2}{w^2}+{\tilde
c}v_\infty \,
 . \label{evl}
\end{equation}
For simplicity wa have also defined a damping length, including both
the
effects of Hubble damping and friction,
\begin{equation}
\frac{1}{\ell_d}=2H+\frac{1}{\ell_f}\, , \label{fric}
\end{equation}
with the friction length scale $\ell_f$ being defined in
\cite{ms1,ms2,thesis}.

Similarly, the velocity equation becomes
\begin{equation}
\frac{dv_\infty}{dt}=\left(1-\frac{v^2_\infty}{w^2}\right)
\left(w^2\frac{k}{L}-\frac{v_\infty}{\ell_d}\right)\, .
\label{evv}
\end{equation}

Note that these are valid for any {\em cosmological}
scenario\footnote{There are some additional subtleties involved
when discussing the mechanism of loop production
in the case of Minkowski space string networks, which make it
quite different from any cosmological scenario. We shall discuss
this important point elsewhere \cite{prep2}.}
We {\em do expect} the loop chopping efficiency ${\tilde{c}}$ to be a
constant,
regardless of the cosmological model, since it is supposed
to be reflecting
a rather deep and fundamental property of the evolution of a
network. Indeed, we think that
whether or not one finds a constant chopping
efficiency can in some sense be seen as a measure of how accurately
the analytic modelling is reproducing the true dynamics of
the network.

We can now re-examine the question of the existence of `scale
invariant'
attractor solutions. Again, scaling solutions of the
form $L\propto t$, $L\propto H^{-1}$
or $L\propto d_H$, together with $v_\infty=const.$ will only exist
provided
one has
\begin{equation}
a(t)\propto t^\beta\, , \qquad \beta=const.\, , \qquad 0<\beta <1\, ,
\label{conda}
\end{equation}
but now we also require
\begin{equation}
\Omega=const.\,
\label{condo}
\end{equation}
The simplest example of the second condition is of course a flat,
$\Omega_{\rm M0}=1$ universe, but there are examples of cosmological
models
which have attractors other than $\Omega=1$ \cite{vsl}.
In any case, note that there
can be additional relations between the values of $\beta$ and $\Omega$
for specific models. Writing $L=\gamma t$ as before, the scaling
solution is now given in the implicit form
\begin{equation}
\gamma^2=w^2\frac{k(k+{\tilde c})}{4\beta(1-\beta)}\, ,
\qquad
v^2=w^2\frac{k(1-\beta)}{\beta(k+{\tilde c})}\, ,
\label{scalvel}
\end{equation}
where $k$ is (implicitly) the constant value of $k(v)$ for the
appropriate
value of velocity,
and
\begin{equation}
w=\frac{2(1-\beta)}{(1-\Omega)\beta k(k+{\tilde
c})}\left[\left(1+\frac{(1-\Omega)\beta k(k+{\tilde c})}{(1-\beta)}\right)^{1/2}-1\right]\, .
\label{scalwww}
\end{equation}
Again, although it may not be immediately obvious, it can be
checked numerically that this solution is well-behaved for all
sensible values
of the parameters. If the two conditions above do not hold,
then a scaling solution will not
exist.

We should also mention another cosmologically important solution:
in an open universe with $\Omega\to 0$, $a\propto t$,
the asymptotic solution is
\begin{equation}
L=At\left(\ln{t}\right)^{1/2}\, \qquad A=\left[\frac{k_{nr}{\tilde
c}}{2(1-k_{nr
})}\right]^{1/2}\sim2.13{{\tilde c}}^{1/2}\, ,
\label{scalopl}
\end{equation}
\begin{equation}
v_\infty=B\left(\ln{t}\right)^{-1/2}\, , \qquad
B=\left[\frac{k_{nr}(1-
k_{nr})}{2{\tilde c}}\right]^{1/2}\sim0.21{{\tilde c}}^{-1/2}\, ,
\label{scalopv}
\end{equation}
with $k_{nr}$ given by(\ref{limsolknr}).
Note that this {\em is not} a scale-invariant solution, since
$H^{-1}=t$ and
$d_H=t\ln{t}$. In other words, by looking at the network one would be
able
to determine when the curvature-dominated period had started.

\section{Discussion and conclusions}
\label{sdsc}

In this paper we have presented a modified version of the
velocity-dependent
one-scale (VOS) model \cite{ms1,ms2,ms3,thesis} which depends
on a single free parameter, the loop
chopping efficiency ${\tilde c}$. We have tested it against the
largest and most accurate numerical simulations to date
\cite{frac,prep2},
and we find that it provides a good fit to the large-scale
scaling properties of the string network in both the radiation and
matter
epochs, as well as in the transition between the two eras---we will
describe these tests elsewhere \cite{test}.
These facts and its intrinsic simplicity make this model particularly
suited
for any analytic or semi-analytic study of cosmic strings where one is
only
interested in the large-scale features of the network.

We have re-analysed some simple evolutionary properties of
cosmic string networks in the light of the VOS model and corresponding
numerical simulations. An important conclusion to note is that any
realistic
cosmic string network is {\it not} scaling at any time from just
before the 
epoch of equal matter and radiation through to the present day.
This is something that must be properly taken into account
particularly when 
discussing
string-seeded structure formation scenarios with GUT-scale strings. 
The extended VOS model is also valid when deviations from scaling are
even 
larger at late times in 
a universe which becomes dominated by curvature or a cosmological
constant.

Finally, we considered the effects of radiation back-reaction
on the scaling properties of the long string network, and we have
shown that although
the existence (or otherwise) of a scale-invariant attractor solution
will
not be affected, the quantitative scaling properties can be.
In some cases, the suppression of string density can be quite dramatic
(as 
we saw for small-scale global string simulations), although the string
velocities 
always remain relativistic.  For the most part, however, the density
of a cosmic
string network, whether local or global is only affected slightly 
by radiation backreaction effects.
 
Despite the many virtues of the VOS model, we are  aware, of course, 
that the small number of available degrees of freedom means that
this model is unable to provide a proper description of the
small-scale
properties of the network; these are important in a number of
cosmological
scenarios (and sometimes even crucial). Nevertheless, we believe that 
the phenomenological parameter $k$ does 
encode some important small-scale structure effects, though clearly 
a more detailed analytic and numerical study is still required.
A number of possible approaches to
the problem of string small-scale structure 
have been suggested in the literature \cite{emb,ack}, and our own 
analysis using Carter's elastic string model \cite{carter} will be
discussed 
in a forthcoming publication \cite{prep3}.

%%%%%%%%%%%%%%%%%%%%%%%%%%%%%%%%%%%%%%%%%%%%%%%%%%%%%%%%%%%%%%%%%%%%%%%%%%
\acknowledgements

We would like to thank Pedro Avelino, Brandon Carter, Jonathan Moore,
Levon Pogosian, Tanmay Vachaspati and Proty
Wu for useful conversations. C.M. also acknowledges discussions with a
number
of participants in the EC Summer School `Multi-fractals---Mathematics
and
Applications', held at the Isaac Newton Institute. C.M. is funded by
FCT
(Portugal) under `Programa PRAXIS XXI' (grant no. PRAXIS
XXI/BPD/11769/97).

This work was performed on COSMOS, the Origin2000 owned by the UK
Computational Cosmology Consortium, supported by Silicon Graphics/Cray
Research, HEFCE and PPARC.

%%%%%%%%%%%%%%%%%%%%%%%%%%%%%%%%%%%%%%%%%%%%%%%%%%%%%%%%%%%%%%%%%%%%%%%%%%

%%%%%%%%%%%%%%%%%%%%%%%%%%%%%%%%%%%%%%%%%%%%%%%%%%%%%%%%%%%%%%%%%%%%%%%%%%
%\vfill\eject

\end{document}